\documentclass[english]{IEEEtran}
\usepackage[latin9]{inputenc}
\usepackage{amsbsy}
\usepackage{amssymb}
\usepackage{graphicx}
\usepackage{babel}
\begin{document}

\title{Precoder Design for Orthogonal Space-Time Block Coding based Cognitive
Radio with Polarized Antennas}

\author{Abdelwaheb Marzouki, Xin Jin\\
Institut Mines-Telecom, Telecom SudParis, CNRS Samovar UMR 5157, France\\
\{abdelwaheb.marzouki, xin.jin\}@it-sudparis.eu}
\maketitle
\begin{abstract}
The spectrum sharing has recently passed into a mainstream Cognitive
Radio (CR) strategy. We investigate the core issue in this strategy:
interference mitigation at Primary Receiver (PR). We propose a linear
precoder design which aims at alleviating the interference caused
by Secondary User (SU) from the source for Orthogonal Space-Time Block
Coding (OSTBC) based CR. We resort to Minimum Variance (MV) approach
to contrive the precoding matrix at Secondary Transmitter (ST) in
order to maximize the Signal to Noise Ratio (SNR) at Secondary Receiver
(SR) on the premise that the orthogonality of OSTBC is kept, the interference
introduced to Primary Link (PL) by Secondary Link (SL) is maintained
under a tolerable level and the total transmitted power constraint
at ST is satisfied. Moreover, the selection of polarization mode for
SL is incorporated in the precoder design. In order to provide an
analytic solution with low computational cost, we put forward an original
precoder design algorithm which exploits an auxiliary variable to
treat the optimization problem with a mixture of linear and quadratic
constraints. Numerical results demonstrate that our proposed precoder
design enable SR to have an agreeable SNR on the prerequisite that
the interference at PR is maintained below the threshold.\end{abstract}

\section{Introduction}

Cognitive Radio (CR) is an encouraging technology to combat the spectrum
scarcity. In order to further enhance the spectrum utilization, the
spectrum sharing strategy that Primary Users (PUs) and Secondary Users
(SUs) coexist in licensed bands as long as PUs are preserved from
the interference caused by SUs attracts much research efforts. Such
a strategy is tantamount to a multi-user system in which the inter-user
interference mitigation is the core. Various inter-user interference
mitigation techniques for spectrum sharing CR systems have been put
forward. They can be roughly grouped into two categories: power allocation
\cite{Ying-Chang Liang 2011}-\cite{R.-C. Xie2012} and precoding
in Multiple-Input Multiple-Output (MIMO) CR systems \cite{Precoder Optimization in Cognitive Radio with Interference Constraints}-\cite{Linear Precoding for Orthogonal Space-Time Block Coded MIMO-OFDM Cognitive Radio}.

Space Time Block Coding (STBC) exploits time and space diversity in
MIMO systems so as to heighten the reliability of the message signal.
Orthogonal STBC (OSTBC) are contrived in such a fashion that the vectors
of coding matrix are orthogonal in both time and space dimensions.
This feature yields a simple linear decoding at the receiver side
so that no complex matrix manipulation---Singular Value Decomposition
(SVD), for instance, is required for recovering the information bit
from the gathered received symbols. Numerous precoding techniques
have been mooted for unstructured codes. However, these techniques
cannot be applied to OSTBC which should forcibly preserve a special
space-time structure. The precoding design for OSTBC CR systems attracts
less attention in previous work. Such previous work in \cite{Linear Precoding for Orthogonal Space-Time Block Coded MIMO-OFDM Cognitive Radio}
was based on the Maximum Likelihood (ML) space-time decoder, whereas
the ML decoder is a nonlinear method. Inspired by Minimum Variance
(MV) receiver applied for OSTBC multi-access systems \cite{S. Shahbazpanahi 2004}
which used a weight matrix at the receiver side to quell the inter-user
interference, we make use of MV approach to design a precoding matrix
at Secondary Transmitter (ST). 

The precoding matrix at ST is designed to comply with the needs in
our CR system: maximizing the Signal to Noise Ratio (SNR) at Secondary
Receiver (SR) on the premise that the orthogonality of OSTBC is kept,
the interference introduced to Primary Link (PL) by Secondary Link
(SL) is maintained under a tolerable level and the total transmitted
power constraint at ST is satisfied. 

The classic MV beamforming \cite{Robust minimum variance beamforming},
\cite{C. D. Richmond2005} built an optimization problem which includes
only one linear constraint, that cannot administer to the needs in
our CR system. On the other hand, some precoder designs for CR systems
\cite{Spectrum sharing in wireless networks: A QoS-aware secondary multicast approach with worst user performance optimization}
introduced a mixture of linear and quadratic constraints to the optimization
problem which leads to iterative solutions with high computational
complexity. For the purpose of contriving a precoder that applies
to our CR system and provides an analytic solution with low computational
cost, we moot an original precoder design algorithm: we first take
advantage of an optimization problem which includes one linear constraint
with the objective of preserving the orthogonality of OSTBC and making
SL introduce minimal interference to PL for different combinations
of the polarization mode at ST and SR. This optimization problem provides
an analytic solution in terms of an auxiliary variable which is the
system gain on SL. Then we regulate this auxiliary variable using
the quadratic constraints evoked by the transmitted power budget at
ST and the maximum tolerable interference at Primary Receiver (PR).
The polarization mode at ST and SR are conclusively settled on based
upon the maximization criteria of SNR at SR.

The rest of the paper is organized as follows. The system model and
OSTBC are presented in Section II. In Section III, we introduce the
proposed precoder design for OSTBC based CR with polarized antennas.
We report the numerical results and provide insights on the expected
performance in Section IV. Finally, we give the conclusion in Section
V.

\section{System Descriptions}

We consider a CR system that consists of one SL which exploits OSTBC
and one PL. ST and PT are only allowed to communicate with their peers.
ST or PT is equipped with $N_{t}$ antennas and SR or PR is equipped
with $N_{r}$ antennas. The antennas in the same array have identical
polarization mode. On each link, the transmit antenna array or the
receive antenna array is able to switch its polarization mode between
vertical mode $V$ and horizontal mode $H$. We denote by $qt$ and
$qr$, respectively, the transmit antenna array's polarization mode
and the receive antenna array's polarization mode.

\subsection{System Model}

In this paper, we exploit 3GPP Spatial Channel Model (SCM) \cite{3GPP TR }.
The space channel impulse response between a pair of antennas $u$
and $s$ of path $n$ can be expressed as a function in terms of the
polarization channel response and the geometric configuration of the
antennas at both sides of the link:
\begin{equation}
H_{u,s,n}\left(\chi_{BS}^{\left(v\right)},\chi_{BS}^{\left(h\right)},\chi_{MS}^{\left(v\right)},\chi_{MS}^{\left(h\right)},\theta_{n,m,AoD},\theta_{n,m,AoA}\right)\label{eq:3GPPModel}
\end{equation}
where $\chi_{BS}^{\left(v\right)}$ is the BS antenna complex response
for the V-pol component, $\chi_{BS}^{\left(h\right)}$ is the BS antenna
complex response for the H-pol component, $\chi_{MS}^{\left(v\right)}$
is the MS antenna complex response for the V-pol component, $\chi_{MS}^{\left(h\right)}$
is the MS antenna complex response for the H-pol component, $\theta_{n,m,AoD}$
is the Angle of Departure (AOD) for the $m$th subpath of the $n$th
path and $\theta_{n,m,AoA}$ is the Angle of Arrival (AOA) for the
$m$th subpath of the $n$th path.

We assume that the system is operated over a frequency-flat channel
with $N_{path}$ paths and each path contains only one subpath. For
a point to point communication link, the baseband input-output relationship
at time-slot $t$ is expressed as:

\begin{equation}
\mathbf{y}\left(t\right)=\sqrt{\frac{\rho}{N_{t}}}\mathbf{H}^{qt,qr}\mathbf{x}\left(t\right)+\mathbf{n}\left(t\right)\label{eq:I-R relationship}
\end{equation}
where $\rho$ is the SNR at each receive antenna, $\mathbf{x}\left(t\right)$
is a $N_{t}\times1$ size transmitted signal vector which satisfies
$E\left\{ \mathbf{x}\left(t\right)\mathbf{x^{\mathit{H}}}\left(t\right)\right\} =N_{t}$,
$\mathbf{n}_{j}(t)$ is a $N_{r}\times1$ size complex Gaussian noise
vector at receiver with zero-mean and unit-variance and $\mathbf{H}^{qt,qr}$
is the $N_{r}\times N_{t}$ channel matrix for the specified $qt$
and $qr$ with the entry 
\begin{equation}
H_{u,s}^{qt,qr}=\sum_{n=1}^{N_{path}}H_{u,s,n}\left(\chi_{BS}^{\left(x\neq qt\right)}=0,\,\chi_{MS}^{\left(y\neq qr\right)}=0\right)\label{eq:entry}
\end{equation}
 where $x,y\in\left\{ V,\, H\right\} $. $\mathbf{H}^{qt,qr}$ has
unit variance and satisfies $E\left\{ \mathrm{tr\left(\mathbf{\mathbf{H}^{\mathit{qt,qr}}}\mathbf{\mathbf{H}^{\mathit{qt,qr}}}^{\mathrm{\mathit{H}}}\right)}\right\} =N_{t}N_{r}$.

Assuming that the channel is constant from $t=1$ to $t=T$, then
Equation (\ref{eq:I-R relationship}) can be extended into:

\begin{equation}
\mathbf{Y}=\sqrt{\frac{\rho}{N_{t}}}\mathbf{\mathbf{H}^{\mathit{qt,qr}}}\mathbf{X}+\mathbf{N}\label{eq:extend I-R}
\end{equation}
where $\mathbf{Y}=\left[\mathbf{y}(1),\ldots,\mathbf{y}(T)\right]$,
$\mathbf{X}=\left[\mathbf{x}(1),\ldots,\mathbf{x}(T)\right]$ and
$\mathbf{N}=\left[\mathbf{n}\left(1\right),\ldots,\mathbf{n}\left(T\right)\right]$.

\subsection{Orthogonal Space-Time Block Coding }

If $\mathbf{X}$ is OSTBC matrix, then $\mathbf{X}$ has a linear
representation in terms of complex information symbols prior to space-time
encoding $s_{k},\, k=1,\ldots,K$ \cite{R. G. Cavalcante 2008}:

\begin{equation}
\mathbf{X}=\sum_{k=1}^{K}\left(\mathbf{C}_{k}\mathrm{Re}\left\{ s_{k}\right\} +\mathbf{D}_{k}\mathrm{Im}\left\{ s_{k}\right\} \right)\label{eq:linear representation}
\end{equation}
where $\mathbf{C}_{k}$ and $\mathbf{D}_{k}$ are $N_{t}\times T$
code matrices \cite{M. Gharavi-Alkhansari 2005}.

OSTBC matrix has the following unitary property:

\begin{equation}
\mathbf{X}\mathbf{X}^{H}=\left(\sum_{k=1}^{K}\left|s_{k}\right|^{2}\right)\mathbf{I}_{N_{t}\times N_{t}}\label{eq:OSTBC unitary property}
\end{equation}

In order to represent the relationship between the original symbols
and the received signal by multiplication of matrices, we introduce
the ``underline'' operator \cite{M. Gharavi-Alkhansari 2005} to
rewrite Equation (\ref{eq:I-R relationship}) as:

\begin{equation}
\mathbf{\underline{Y}}=\mathbf{\mathcal{\mathbb{\mathcal{H}}}}^{qt,qr}\mathbf{A}\mathbf{\mathrm{\mathbf{\underline{s}}}}+\underline{\mathbf{N}}\label{eq:underline expression}
\end{equation}
where $\mathbf{s}=\left[s_{1},\ldots,s_{K}\right]$ is the data stream
which is QPSK modulated in this paper, $\mathbb{\mathcal{H}}^{qt,qr}=\left[\begin{array}{cc}
\mathrm{Re}\left\{ \mathbf{I}_{T}\otimes\mathbf{H}^{qt,qr}\right\}  & \mathrm{-Im}\left\{ \mathbf{I}_{T}\otimes\mathbf{H}^{qt,qr}\right\} \\
\mathrm{Im}\left\{ \mathbf{I}_{T}\otimes\mathbf{H}^{qt,qr}\right\}  & \mathrm{Re}\left\{ \mathbf{I}_{T}\otimes\mathbf{H}^{qt,qr}\right\} 
\end{array}\right]$ is the equivalent channel matrix with the specified polarization
mode, $\mathbf{A}=\left[\underline{\mathbf{C}_{1}},\ldots,\underline{\mathbf{C}_{k}},\underline{\mathbf{D}_{1}},\ldots,\underline{\mathbf{D}_{k}}\right]$
is the OSTBC compact dispersion matrix and the ``underline'' operator
for any matrix $\mathbf{P}$ is defined as:

\begin{equation}
\underline{\mathbf{P}}\triangleq\left[\begin{array}{c}
\mathrm{vec}\left\{ \mathrm{Re}\left(\mathbf{P}\right)\right\} \\
\mathrm{vec}\left\{ \mathrm{Im}\left(\mathbf{P}\right)\right\} 
\end{array}\right]\label{eq:underline operator}
\end{equation}
where $\mathrm{vec}\left\{ \bullet\right\} $ is the vectorization
operator stacking all columns of a matrix on top of each other.

The earliest OSTBC scheme which is well known as Alamouti's code was
proposed in \cite{alamouti}. Alamouti's code gives full diversity
in the spatial dimension without data rate loss. The transmission
matrix of Alamouti's code $C_{2}$ is given as:

\begin{equation}
C_{2}=\left[\begin{array}{cc}
s_{1} & s_{2}\\
-s_{2}^{*} & s_{1}^{*}
\end{array}\right]\label{eq:alamouti's code}
\end{equation}

In \cite{V. Tarokh}, Alamouti's code was extended for more antennas.
For instance, four antennas, the transmission matrix of the half rate
code $C_{4}$ is given as:

\begin{equation}
C_{4}=\left[\begin{array}{cccc}
s_{1} & s_{2} & s_{3} & s_{4}\\
-s_{2} & s_{1} & -s_{4} & s_{3}\\
-s_{3} & s_{4} & s_{1} & -s_{2}\\
-s_{4} & -s_{3} & s_{2} & s_{1}\\
s_{1}^{*} & s_{2}^{*} & s_{3}^{*} & s_{4}^{*}\\
-s_{2}^{*} & s_{1}^{*} & -s_{4}^{*} & s_{3}^{*}\\
-s_{3}^{*} & s_{4}^{*} & s_{1}^{*} & -s_{2}^{*}\\
-s_{4}^{*} & -s_{3}^{*} & s_{2}^{*} & s_{1}^{*}
\end{array}\right]\label{eq:C4}
\end{equation}

\section{Precoder for OSTBC based CR with Polarized Antennas}

We design a precoding matrix at ST which acts on the entry of the
OSTBC compact dispersion matrix and has no influence on the codes'
structure. Our precoder design relies on the equivalent transmit correlation
matrix on the link between ST and PR (SPL). This matrix can be estimated
easily by SU in the sensing step and enables our precoder design to
regulate the interference introduced by SL to PL.

\subsection{Constraints from SL}

With the precoding operation, the received signal at SR for the specified
polarization mode at ST and SR can be expressed as:

\begin{equation}
\underline{\mathbf{Y_{\mathit{ST,SR}}^{\mathit{qt,qr}}}}=\sqrt{\frac{\rho_{SR}}{N_{t}}}\mathbb{\mathcal{H}}_{ST,SR}^{qt,qr}\mathbf{W}^{qt,qr}\mathbf{A}\mathbf{\underline{s}}+\underline{\mathbf{N}}\label{eq:received signal}
\end{equation}
where $\rho_{SR}$ is the SNR at each receive antenna of SR, $\mathbb{\mathcal{H}}_{ST,SR}^{qt,qr}$
is the SL equivalent channel matrix with the specified polarization
mode at ST and SR, $\mathbf{\mathbf{W}^{\mathit{qt,qr}}}$ is the
precoding matrix for the specified polarization mode at ST and SR. 

A straightforward approach to estimate the transmitted signal from
ST is using the following soft output detector:

\begin{eqnarray}
\mathbf{\underline{\hat{s}}} & = & \mathbf{A}^{T}\mathbb{\mathcal{H}}_{ST,SR}^{qt,qr^{T}}\underline{\mathbf{Y_{\mathit{ST,SR}}^{\mathit{qt,qr}}}}\label{eq:soft detector}\\
 & = & \sqrt{\frac{\rho_{SR}}{N_{t}}}\mathbf{A}^{T}\mathbb{\mathcal{H}}_{ST,SR}^{qt,qr^{T}}\mathbb{\mathcal{H}}_{ST,SR}^{qt,qr}\mathbf{\mathbf{W}^{\mathit{qt,qr}}}\mathbf{A}\mathbf{\underline{s}}+\mathbf{A}^{T}\mathbb{\mathcal{H}}_{ST,SR}^{qt,qr^{T}}\underline{\mathbf{N}}\nonumber 
\end{eqnarray}

The OSTBC structure conservation puts forward the following constraint:

\begin{equation}
\mathbf{A}^{T}\mathbb{\mathcal{H}}_{ST,SR}^{qt,qr^{T}}\mathbb{\mathcal{H}}_{ST,SR}^{qt,qr}\mathbf{\mathbf{W}^{\mathit{qt,qr}}}\mathbf{A}=\alpha^{qt,qr}\mathbf{I}_{2K}\label{eq:structure constraint}
\end{equation}
where $\alpha^{qt,qr}$ is the system gain on SL for the specified
polarization mode at ST and SR which will be adjusted to satisfy the
other constraints.

Additionally, the transmitted power budget at ST induces another constraint:

\begin{equation}
P_{t}^{qt,qr}\leq P_{tmax}\label{eq:power constraint}
\end{equation}
where $P_{t}^{qt,qr}=\frac{\rho_{SR}}{N_{t}}\mathrm{tr}\left(\mathbf{W^{\mathit{qt,qr}}}^{T}\mathbf{W}^{qt,qr}\right)$
and $P_{tmax}$ are, respectively, the transmitted power for the specified
polarization mode at ST and SR and the maximum transmitted power at
ST.

\subsection{Constraints from PL}

The received signal at PR from ST is deemed as baleful signal by PL
and can be expressed as:

\begin{equation}
\underline{\mathbf{Y_{\mathit{ST,PR}}^{\mathit{qt,qr'}}}}=\sqrt{\frac{\rho_{PR}}{N_{t}}}\mathbb{\mathcal{H}}_{ST,PR}^{qt,qr'}\mathbf{W}^{qt,qr}\mathbf{A}\mathbf{\underline{s}}+\underline{\mathbf{N}}\label{eq:received signal PR ST}
\end{equation}
where $\rho_{PR}$ is the SNR at each receive antenna of PR and $\mathbb{\mathcal{H}}_{ST,PR}^{qt,qr'}$
is the equivalent channel matrix for the specified polarization mode
at ST and PR.

The interference power introduced by SL to PL for the specified polarization
mode at ST and PR can be calculated as: 

\begin{eqnarray}
P_{ST,PR}^{qt,qr'} & = & \mathrm{tr}\left[E\left(\mathbf{\underline{\mathbf{\mathbf{Y_{\mathit{ST,PR}}^{\mathit{qt,qr'}}}}}}\,\mathbf{\underline{\mathbf{\mathbf{Y_{\mathit{ST,PR}}^{\mathit{qt,qr'}}}}}}^{H}\right)\right]\label{eq:Pinterf}\\
 & = & \frac{\rho_{SR}}{N_{t}}\mathrm{tr}\left(\mathbf{W^{\mathit{qt,qr}}}^{T}R_{\mathit{PR,ST}}^{\mathit{qt,qr'}}\mathbf{W}^{qt,qr}\right)\nonumber 
\end{eqnarray}
where $\mathcal{R_{\mathit{PR,ST}}^{\mathit{qt,qr'}}}=E\left(\mathbb{\mathcal{H}}_{PR,ST}^{qt,qr'T}\mathbb{\mathcal{H}}_{PR,ST}^{qt,qr'*}\right)$
is the equivalent transmit correlation matrix on SPL for the specified
polarization mode at ST and PR. The maximum tolerable interference
power $\eta$ at PR evokes the following constraint:

\begin{equation}
P_{ST,PR}^{qt,qr'}\leq\eta\label{eq:interf power cons}
\end{equation}

\subsection{Minimum Variance Algorithm}

SU can dominate the configuration of the precoding matrix and the
polarization mode on SL, while SU has no eligibility to select the
polarization mode on PL. Our algorithm is based on an optimization
problem which includes one linear constraint with the objective of
preserving the orthogonality of OSTBC and making SL introduce minimal
interference to PL for different combinations of the polarization
mode at ST and SR. This optimization problem provides an analytic
solution in terms of an auxiliary variable which is the system gain
on SL. Then this auxiliary variable is regulated by using the quadratic
constraints evoked by the transmitted power budget at ST and the maximum
tolerable interference at PR. The polarization mode at ST and SR are
conclusively settled on based upon the maximization criteria of SNR
at SR.

Such an optimization problem that includes one linear constraint is
described as follow: 

\begin{equation}
\left(\widehat{\mathbf{W^{\mathit{qt,qr}}}},\,\widehat{qt},\,\widehat{qr}\right)=\arg\min_{\mathbf{W^{\mathit{qt,qr}}},\, qt,\, qr}\frac{\rho_{SR}}{N_{t}}\mathrm{tr}\left(\mathbf{W^{\mathit{qt,qr}}}^{T}\mathcal{R_{\mathit{PR,ST}}^{\mathit{qt,qr'}}}\mathbf{W^{\mathit{qt,qr}}}\right)\label{eq:minimum}
\end{equation}

\begin{equation}
\mathrm{subject}\;\mathrm{to}:\;\mathrm{tr}\left(\mathbf{A}^{T}\mathbb{\mathcal{H}}_{ST,SR}^{qt,qr^{T}}\mathbb{\mathcal{H}}_{ST,SR}^{qt,qr}\mathbf{\mathbf{W^{\mathit{qt,qr}}}}\mathbf{A}\mathit{\mathrm{-}}\alpha^{qt,qr}\mathbf{I}_{2K}\right)=0\label{eq:tr}
\end{equation}

We exploit the method of Lagrange multipliers to find $\widehat{\mathbf{W^{\mathit{qt,qr}}}}$
for each combination of the polarization mode at ST and SR. The Lagrangian
function can be written as: 

\begin{eqnarray}
L\left(\mathbf{\mathbf{W^{\mathit{qt,qr}}}},\,\mathbf{\mathbf{\boldsymbol{\Lambda}}}\right) & = & \frac{\rho_{SR}}{N_{t}}\mathrm{tr}\left(\mathbf{\mathbf{W^{\mathit{qt,qr}}}}^{T}\mathcal{R_{\mathit{PR,ST}}^{\mathit{qt,qr'}}}\mathbf{\mathbf{W^{\mathit{qt,qr}}}}\right)\nonumber \\
 &  & -\mathrm{tr}\left(\mathbf{\mathbf{\boldsymbol{\Lambda}}}^{T}\left(\mathbf{A}^{T}\mathcal{R_{\mathrm{\mathit{ST,SR}}}^{\mathit{qt,qr}}}\mathbf{\mathbf{W^{\mathit{qt,qr}}}}\mathbf{A}\mathit{\mathrm{-}}\alpha^{qt,qr}\mathbf{I}_{2K}\right)\right)\nonumber \\
\label{eq:LR}
\end{eqnarray}
where $\mathcal{R_{\mathrm{\mathit{ST,SR}}}^{\mathit{qt,qr}}}=\mathbb{\mathcal{H}}_{ST,SR}^{qt,qr^{T}}\mathbb{\mathcal{H}}_{ST,SR}^{qt,qr}$
and $\boldsymbol{\Lambda}$ is a $2K\times2K$ size matrix of Lagrange
multipliers. 

By differentiating the Lagrange function with respect to $\mathbf{\mathbf{W^{\mathit{qt,qr}}}}$
and equating it to zero, we obtain an analytic solution in terms of
$\alpha^{qt,qr}$ which is expressed as:

\begin{equation}
\widehat{\mathbf{W^{\mathit{qt,qr}}}}=\alpha^{qt,qr}\mathcal{R_{\mathit{PR,ST}}^{\mathit{qt,qr'}}}^{-1}\mathcal{R_{\mathrm{\mathit{ST,SR}}}^{\mathit{qt,qr}}}\mathbf{A}\mathbf{Q^{\mathit{qt,qr}}}\mathbf{A}^{T}\label{eq:analytical solution}
\end{equation}
where $\mathbf{Q^{\mathit{qt,qr}}}=\left(\mathbf{A}^{T}\mathcal{R_{\mathrm{\mathit{ST,SR}}}^{\mathit{qt,qr}}}\left(R_{\mathit{PR,ST}}^{\mathit{qt,qr'}}\right)^{-1}\mathbf{A}\right)^{-1}$. 

The estimated interference power at PR can be expressed in terms of
$\alpha^{qt,qr}$ as: 
\begin{equation}
\widehat{P_{ST,PR}^{qt,qr}}=\frac{\rho_{SR}\left(\alpha^{qt,qr}\right)^{2}\mathrm{tr}\left(\mathbf{Q^{\mathit{qt,qr}}}\right)}{N_{t}}\label{eq:interfpower}
\end{equation}

The estimated SNR at SR can be written in terms of $\alpha^{qt,qr}$
as: 

\begin{equation}
\widehat{SNR_{ST,PR}^{qt,qr}}=\frac{\rho_{SR}\left(\alpha^{qt,qr}\right)^{2}\gamma^{qt,qr}}{N_{t}}\label{eq:SNR}
\end{equation}
where 
\begin{eqnarray}
\gamma^{qt,qr}=\nonumber \\
\mathrm{tr}\left(\mathbf{Q^{\mathit{qt,qr}}}\mathbf{A}^{T}\left(\mathcal{R_{\mathrm{\mathit{ST,SR}}}^{\mathit{qt,qr}}}R_{\mathit{PR,ST}}^{\mathit{qt,qr'}}{}^{-1}\right)^{2}\mathcal{R_{\mathrm{\mathit{ST,SR}}}^{\mathit{qt,qr}}}\mathbf{AQ^{\mathit{qt,qr}}}\right)\label{eq:v}
\end{eqnarray}
.

The estimated transmit power at ST in terms of $\alpha^{qt,qr}$ is
given by:

\begin{equation}
\widehat{P_{t}^{qt,qr}}=\frac{\rho_{SR}\left(\alpha^{qt,qr}\right)^{2}\delta^{qt,qr}}{N_{t}}\label{eq:TR POWER}
\end{equation}
where 

\begin{equation}
\delta^{qt,qr}=\mathrm{tr}\left(\mathbf{Q}^{qt,qr}\mathbf{A}^{T}\mathcal{R_{\mathrm{\mathit{ST,SR}}}^{\mathit{qt,qr}}}\left(R_{\mathit{PR,ST}}^{\mathit{qt,qr'}}\right)^{-2}\mathcal{R_{\mathrm{\mathit{ST,SR}}}^{\mathit{qt,qr}}}\mathbf{AQ}^{qt,qr}\right)\label{eq:o}
\end{equation}

We derive $\alpha^{qt,qr}$ by substituting $\widehat{P_{t}^{qt,qr}}$
and $\widehat{P_{ST,PR}^{qt,qr}}$ into Equation (\ref{eq:power constraint})
and Equation (\ref{eq:interf power cons}) which indicate the transmitted
power budget constraint and the maximum tolerable interference constraint:

\begin{equation}
\alpha^{qt,qr}=\min\left(\sqrt{\frac{N_{t}}{\delta^{qt,qr}}},\,\sqrt{\frac{N_{t}\eta}{\rho_{SR}\mathrm{tr}\left(\mathbf{Q^{\mathit{qt,qr}}}\right)}}\right)\label{eq:a}
\end{equation}
Therefore the estimated SNR at SR can be determined as:

\begin{equation}
\widehat{SNR_{ST,PR}^{qt,qr}}=\min\left(\frac{\rho_{SR}}{\delta^{qt,qr}},\frac{\eta}{\mathrm{tr}\left(\mathbf{Q}^{\mathit{qt,qr}}\right)}\right)\gamma^{qt,qr}\label{eq:SNR-1}
\end{equation}

Based upon the maximization criteria of SNR at SR, Finally, we destine
the estimated polarization mode of ST and SR as:

\begin{equation}
\left(\widehat{qt},\widehat{qr}\right)=\arg\max_{qt,qr}\left[\min\left(\frac{\rho_{SR}}{\delta^{qt,qr}},\frac{\eta}{\mathrm{tr}\left(\mathbf{Q}^{qt,qr}\right)}\right)\gamma^{qt,qr}\right]\label{eq:qt qr}
\end{equation}

\section{Numerical Results}

For the purpose of validating our proposed precoding design algorithm,
we simulated our CR system using the proposed precoder design algorithm
and measure the SNR at SR by using varying maximum transmitted power
at ST and a reasonable interference threshold at PR.

We firstly carried out our simulation with Alamouti's code at ST for
different combinations of $qt$ and $qr$ on SL under different multipath
scenarios. Then, we executed our simulation with different codes for
different number of transmit antennas at ST based upon a determinate
combination of $qt$ and $qr$ on SL and multipath scenario. In both
simulation scenarios, the Signal to Interference plus Noise Ratio
(SINR) threshold to perceive the received signal at PR was chosen
equal to $0\mathrm{dB}$ and the Cross-polar Discrimination (XPD)
was set to $8\mathrm{dB}$. The channel matrix on each link was modeled
according to 3GPP SCM. Since the status of polarization at PR $qr'$
is normally unidentified for SU, the equivalent transmit correlation
matrix on SPL becomes random. This thereby results in a random SNR
at SR. In our simulation, we calculated the SNR at SR in terms of
the polarization tilt angle at PR by introducing a rotation matrix
to the equivalent transmit correlation matrix on SPL. We assumed that
the polarization tilt angle at PR follows a continuous uniform distribution
between $0$ and $\frac{\pi}{2}$. Then we sampled uniformly over
the range of the polarization tilt angle at PR and calculated the
SNR at SR for each sample of tilt angle. Finally, we worked out an
average the SNR at SR to evaluate the system performance.

\subsection{Performance Analysis of Polarization Diversity }

\begin{figure}
\includegraphics[width=9cm,height=6cm]{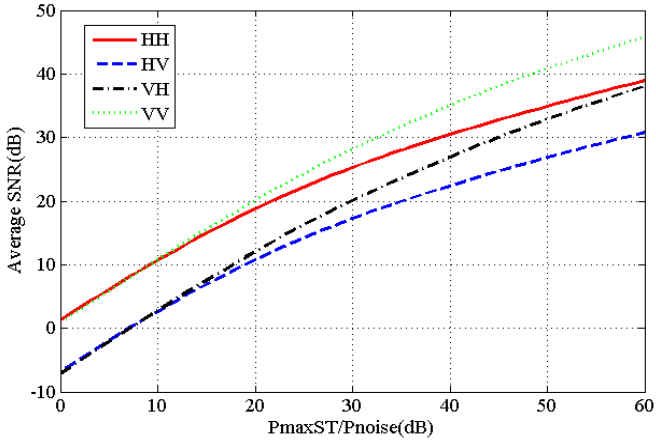}

\caption{Average SNR at SR versus $P_{maxSU}/P_{noise}$ for different polarization
mode on SL under single path scenario\label{fig:fig1}}
\end{figure}

\begin{figure}
\includegraphics[width=9cm,height=6cm]{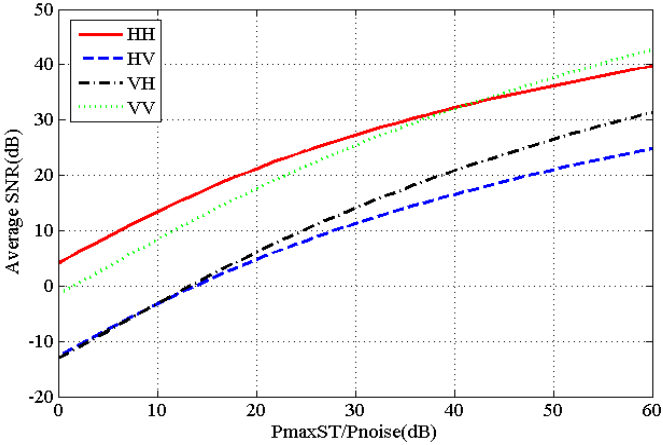}

\caption{Average SNR at SR versus $P_{maxSU}/P_{noise}$ for different polarization
mode under four paths scenario\label{fig:fig2}}
\end{figure}

We simulated a CR system, where ST is equipped with 2 antennas, SR
is equipped with 1 antenna and PR is equipped with 2 antennas. We
observe the variation of the average SNR at SR for different combinations
of $qt$ and $qr$ on SL as the transmit power at ST increases. First,
we set SL channel as a 2-path frequency flat fading channel and SPL
channel as a single path frequency flat fading channel. The variation
tendencies in this scenario were depicted in Fig.\ref{fig:fig1}.
Then we reset SPL channel as a 4-path frequency flat fading channel
and the corresponding variation tendencies were shown in Fig.\ref{fig:fig2}.
The average SNR at SR for a large number of samples leads to the smooth
curves. As the transmit power at ST increases, the average SNR at
SR of all different combinations of $qt$ and $qr$ on SL exhibit
uptrend in both scenarios and linear increase is obtained when $P_{maxSU}/P_{noise}$
are below $15\mathrm{dB}$ in both scenarios, where $P_{noise}$ denotes
the noise power at SR. The mismatch of $qt$ and $qr$ on SL induces
a $15\mathrm{dB}$ gap between the matched modes and the mismatched
modes when the average SNR at SR has linear increase in the first
scenario. When we enhanced the number of paths in SPL channel, the
average SNR at SR for the mismatched modes was declined by $6\mathrm{dB}$
and the gap was enlarged in the second scenario. 

\subsection{Performance Analysis of Transmit Antennas Diversity}

In the second simulation, we aimed to observe the average SNR at SR
by using different number of transmit antennas. In the first circumstance,
2 transmit antennas and Alamouti's code $C_{2}$ were utilized at
ST. In the second circumstance, 4 transmit antennas and the half rate
code $C_{4}$ were utilized at ST. In both circumstances, we set $qt=V$
and $qr=V$. SR is equipped with 1 antenna and PR is equipped with
4 antennas. The number of paths is chosen equal to 2 on SL and 6 on
SPL. 

For the case of 2 transmit antennas at ST, the SNR at SR reaches the
saturation point at $20\mathrm{dB}$ when $P_{maxSU}/P_{noise}$ achieves
$40\mathrm{dB}$. Compare to the previous results in Fig. \ref{fig:fig1}
and \ref{fig:fig2}, the SNR at SR reaches the saturation point faster
due to the increase in number of paths on the SPL. However, the increase
in number of antennas will significantly delay the arrival of the
saturation point even the number of paths on the SPL is also increased.
For the case of 4 transmit antennas at ST, the SNR at SR reaches the
saturation point at $65\mathrm{dB}$ when $P_{maxSU}/P_{noise}$ achieves
$100\mathrm{dB}$.

\begin{figure}
\includegraphics[width=9cm,height=6cm]{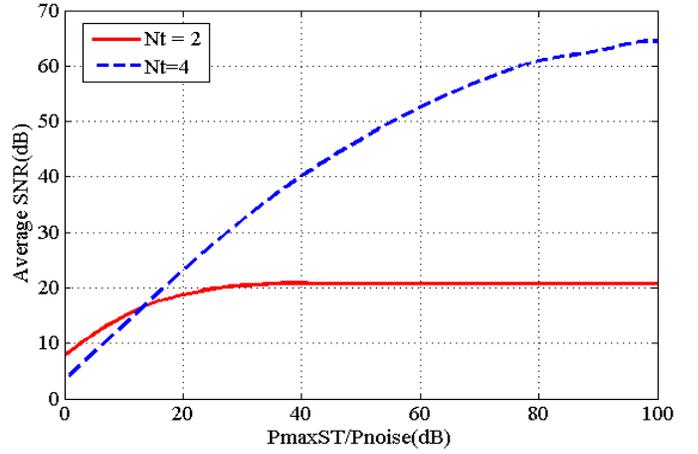}\caption{Average SNR at SR versus $P_{maxSU}/P_{noise}$ for different number
of transmit antennas at ST\label{fig:fig3}}
\end{figure}

\section{Conclusions}

A linear precoder design which aims at alleviating the interference
at PR for OSTBC based CR has been introduced. One of the principal
contributions is to endow the conventional prefiltering technique
with the excellent features of OSTBC in the context of CR. The prefiltering
technique has been optimized for the purpose of maximizing the SNR
at SR on the premise that the orthogonality of OSTBC is kept, the
interference introduced to PL by SL is maintained under a tolerable
level and the total transmitted power constraint is satisfied. Numeral
Results have shown that polarization diversity contributes to achieve
better SNR at SR, moreover, the increase in number of antennas will
significantly delay the arrival of the saturation point for the SNR
at SR.

\section*{Acknowledgements} This research is supported by SACRA project (FP7-ICT-2007-1.1, European Commission-249060).

\end{document}